\documentclass[fleqn,usenatbib]{mnras}
\usepackage{newtxtext,newtxmath}
\usepackage[T1]{fontenc}

\DeclareRobustCommand{\VAN}[3]{#2}
\let\VANthebibliography\thebibliography
\def\thebibliography{\DeclareRobustCommand{\VAN}[3]{##3}\VANthebibliography}

\usepackage{graphicx}
\usepackage{amsmath}

\title[CH$_2$DOH Starless Core Survey]{A Survey of CH$_2$DOH Towards Starless and Prestellar Cores in the Taurus Molecular Cloud}

\author[H. E. Ambrose et al.]{
Hannah E. Ambrose,$^{1}$\thanks{E-mail: heha2455@email.arizona.edu (HEA) \newline Senior Honors Thesis at The University of Arizona}
Yancy L. Shirley,$^{1}$\thanks{E-mail: yshirley@email.arizona.edu (YLS)}
Samantha Scibelli$^{1}$
\\
$^{1}$Steward Observatory, The University of Arizona, 933 N. Cherry Ave., Tucson, AZ 85721
}

\pubyear{2020}

\begin{document}
\label{firstpage}
\pagerange{\pageref{firstpage}--\pageref{lastpage}}
\maketitle

\begin{abstract}
Recent observations indicate that organic molecules are prevalent towards starless and prestellar cores. Deuteration of these molecules has not been well-studied during the starless phase. Published observations of singly-deuterated methanol, CH$_2$DOH, have only been observed in a couple of well-studied, dense and evolved prestellar cores (e.g. L1544, L183). Since the formation of gas-phase methanol during this cold phase is believed to occur via desorption from the icy grain surfaces, observations of CH$_2$DOH may be useful as a probe of the deuterium fraction in the ice mantles of dust grains. We present a systematic survey of CH$_2$DOH towards 12 starless and prestellar cores in the B10 region of the Taurus Molecular Cloud.  Nine of the twelve cores are detected with  [CH$_2$DOH]/[CH$_3$OH] ranging from $< 0.04$ to 0.23$^{+0.12}_{-0.06}$ with a median value of $0.11$.  Sources not detected tend to have larger virial parameters and larger methanol linewidths than detected sources.
The results of this survey indicate  that deuterium fractionation of organic molecules, such as methanol, during the starless phase may be  more  easily detectable than previously thought.
\end{abstract}


\section{Introduction}

Starless cores are dense ($> 10^4$ cm$^{_3}$) clumps of gas and dust, typically with temperatures $\sim$10 K \citep{1989ApJS...71...89B, 2007prpl.conf...17D, 2007ARA&A..45..339B}. When a starless core becomes gravitationally-bound, it enters the stage of star formation immediately prior to the formation of a protostar called a prestellar core \citep{2014prpl.conf...27A}.  Most prestellar cores are low-mass and coalesce into stars of a few solar masses or less.  Starless and especially prestellar cores are relatively calm, non-turbulent environments which lack disruptive features such as internal heat sources, strong shocks and outflows, and strong temperature gradients,
(\citealt{2001ApJ...557..193E}, \citealt{2016IAUS..315...95T}); therefore, 
they are a useful location within which to study the initial chemical conditions of
star formation and ultimately, planet formation.

In the past decade, observations have shown that complex organic molecules (molecules with more than 5 atoms containing C, H, and either N or O; \citealt{2009ARA&A..47..427H}) are detectable towards prestellar cores (e.g., \citealt{2012A&A...541L..12B}, \citealt{2016ApJ...830L...6J}, \citealt{2018ApJ...854..116S}, \citealt{2020ApJ...891...73S}).  A recent survey of 31 starless and prestellar cores in the L1495--B218 regions of Taurus showed methanol (100 percent detection rate) and acetaldehyde (70 percent detection rate) to be extremely common in starless and prestellar cores \citep{2020ApJ...891...73S}. Given the typical phase lifetimes of starless cores with densities of $10^5$ cm$^{-3}$ \citep{2014prpl.conf...27A}, these results indicate complex organic molecule formation is occurring at least hundreds of thousands of years prior to the formation of a protostar.

The deuteration of the complex organic molecules -- the replacement of one or more hydrogen atoms with deuterium -- is one aspect of complex organic molecule chemistry that has not been well studied to date during the starless core phase.  Deuterated COMs are readily observed during low-mass protostellar phases  \citep{2002P&SS...50.1205L,2002A&A...393L..49P,2006A&A...453..949P,2017A&A...606L...7B, 2018A&A...620A.170J, 2019A&A...623A..69M, 2019A&A...632A..19T, 2020A&A...635A..48M, 2020arXiv200506784V} 
but there are few observations towards prestellar cores (cf. \citealt{2014A&A...569A..27B, 2019A&A...622A.141C}).  Deuterium is primarily formed in the first few minutes of the Big Bang \citep{1976Natur.263..198E} and is later 
incorporated into $^3$He and heavier elements within stellar interiors. While the elemental D$/$H ratio is $\sim 2.6 \times 10^{-5}$, \citep{2016ApJ...830..148C,2016A&A...594A..13P}, in cold, dense environments such as starless and prestellar cores, molecular deuteration ratios can reach values several orders of magnitude above the cosmic abundance \citep{2014prpl.conf..859C}.
 
The fractionation of many commonly observed deuterated molecules (i.e. DCO$^+$, N$_2$D$^+$, NH$_2$D, etc.) is thought to proceed through gas phase reactions with H$_2$D$^+$, D$_2$H$^+$, and D$_3^+$ \citep{1989ApJ...340..906M,2003ApJ...591L..41R, 2019RSPTA.37780401C}.  For instance, NH$_3$ can react with H$_2$D$^+$ to form the intermediate ion NH$_3$D$^+$ which then can dissociatively recombine with an electron into NH$_2$D or NH$_3$ (with a branching ratio).  This mechanism is unlikely to work for CH$_3$OH because the reaction of CH$_3$OH$_2^+$ $+ \,\mathrm{e}^-$ has been measured in the laboratory and dissociatively recombines into CH$_3$OH only a small fraction of the time \citep{2006FaDi..133..177G}. Presumably, the same result is true for the deuterated ion CH$_2$DOH$_2^+$.  Thus, significant gas phase deuterium fractionation of methanol in cold ($\sim$10 K) environments appears unlikely.

CH$_2$DOH is a strong candidate for the study of fractionation on the icy surfaces of dust grains.  Models suggest that at least some methanol fractionation is occurring within grain ices, as the gas phase abundance of methanol cannot be explained by pure gas phase ion-molecule chemistry
(\citealt{2006FaDi..133..177G}, shown experimentally).
Instead, CH$_3$OH is thought to form on the icy surfaces of dust grains through the successive hydrogenation of CO,
i.e. CO + H $\rightarrow$ HCO + H $\rightarrow$ H$_2$CO + H $\rightarrow$ CH$_2$OH + H $\rightarrow$ CH$_3$OH \citep{1982A&A...114..245T}. Singly-deuterated methanol is thought to form according to the same process, with one of the H atoms being replaced by deuterium while on the grain \citep{1983A&A...119..177T}. 
In fact, these formation methods have been proven experimentally for both methanol \citep{2002ApJ...571L.173W} and deuterated methanol \citep{2007msl..confE..52W}.

Exactly how methanol and deuterated methanol enter the gas phase is still a matter of discussion. Proposed non-thermal mechanisms for COMs to leave ice mantles in cold, prestellar environments include UV-induced photo-desorption, UV-induced codesorption, local heating or sputtering by cosmic rays, and reactive desorption, with each of these mechanisms facing varying degrees of theoretical and laboratory scrutiny (see
\citealt{2018ApJ...853..102C},
\citealt{2020ApJ...897...56P}, and references therein).
While the relative importance of these desorption mechanisms is still not known, observations of CH$_2$DOH in the gas phase through its millimeter-wavelength rotational transitions provide a link to studying deuteration on the icy surfaces of dust grains, and provide observational constraints on chemical models during the starless and prestellar core phases.

\begin{figure}
    \centering
    \includegraphics[width=84mm]{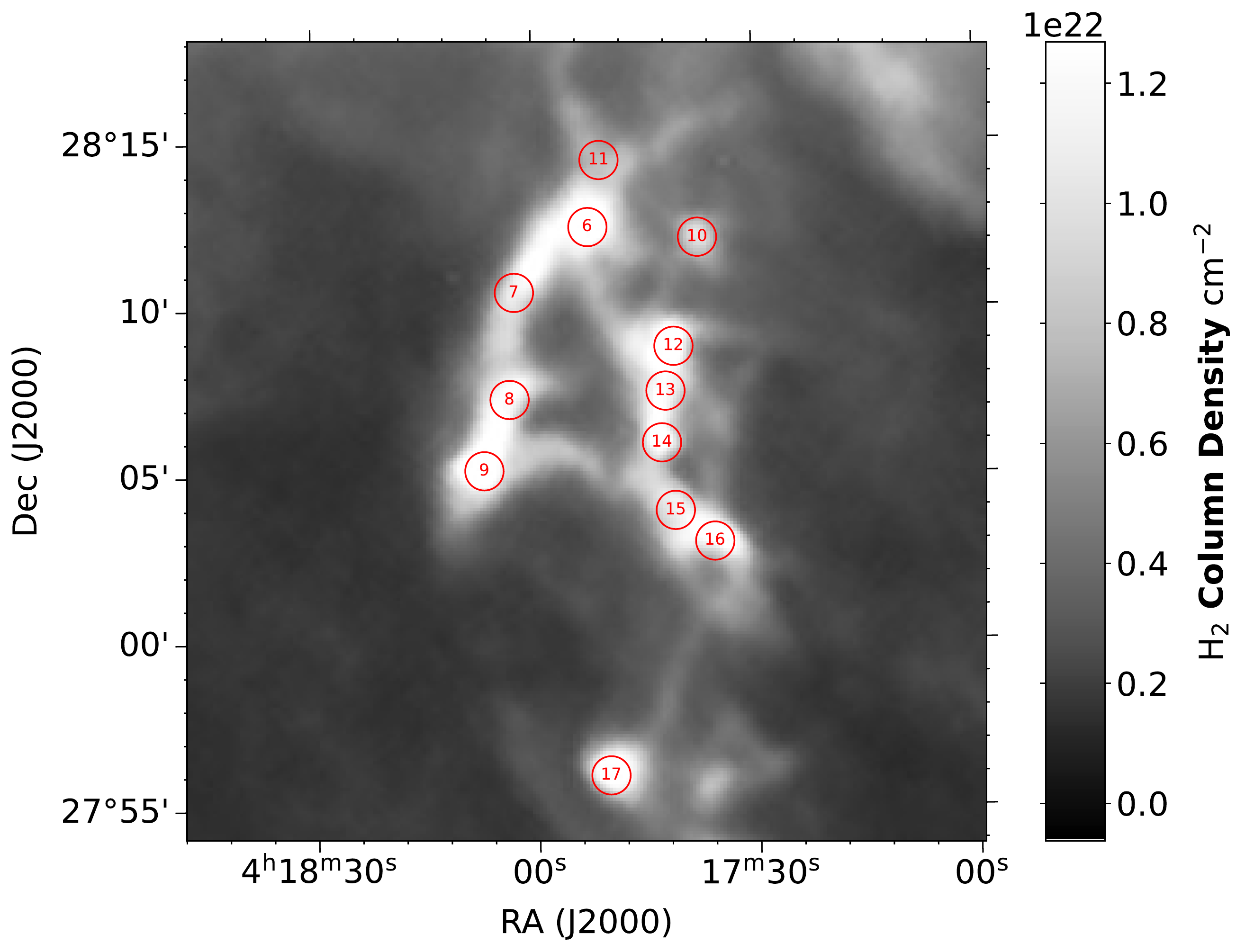}
    \caption{The B10 region of Taurus.  Greyscale image is H$_2$ column density derived from \textit{Herschel Space Observatory} observations \citep{2013A&A...550A..38P, 2016MNRAS.459..342M}. The positions of the NH$_3$-identifed cores \citep{2015ApJ...805..185S} are indicated by circles with the beamsize of the 12 m. }
 
    \label{fig:b10}
\end{figure}

For this survey, we observed the complete population of starless and prestellar dense cores within a single region of a molecular cloud for the presence of singly-deuterated methanol, CH$_2$DOH. In doing so, we look to understand the prevalence and abundance of the molecule in the incipient phases of star formation, as well as to explore deuterium fractionation within the surface ices of dust grains. This survey targeted low-mass (0.31--1.22 M$_{\odot}$) starless and prestellar dense cores in the Barnard 10 (B10) region of the Taurus Molecular Cloud, which were identified in the NH$_3$ mapping survey of \cite{2015ApJ...805..185S}. The twelve starless and prestellar cores, Seo06--Seo17, contain the highest average abundance of methanol among the L1495--B218 regions \citep{2020ApJ...891...73S}.
The B10 region has no embedded protostars (no Class 0/I/II protostars) \citep{2010ApJS..186..259R} and is considered to be a less-evolved, quiescent region free from stellar and protostellar activity \citep{2013A&A...554A..55H, 2015ApJ...805..185S}.
Figure\,\ref{fig:b10} provides a map of these cores within the B10 region.

We present observations of CH$_2$DOH (\S2) and calculate the deuterium fraction (\S3) which is compared to models of grain surface deuteration as well as evolutionary indicators of the cores (\S4).

\section{Observations}

Observations were taken with the Arizona Radio Observatory (ARO) 12 m Radio Telescope on Kitt Peak for 16 shifts between 2018 June and 2018 December towards the NH$_3$ peak positions (see Table \ref{physparams}) from \cite{2015ApJ...805..185S}. We first tuned to the J$_{\rm{K_a,K_c}} =$ 2$_{0,2}$ e$_0$--1$_{0,1}$ e$_0$ (a-type) transition of CH$_2$DOH at 89.407817 GHz using the 3 mm receiver (ALMA mixers). Seo06 and 12 were observed in 2018 June using Millimeter AutoCorrelator (MAC) spectrometer. The observations have 0.0819 km s$^{-1}$ (24.4 kHz) resolution and a FWHM beamsize of 67.6". We observed the remainder of the B10 region for the 89 GHz transition from 2018 October to 2018 November with the new AROWS spectrometer, which had become available in  October. The AROWS spectrometer offered an improved resolution of 0.131 km s$^{-1}$ ($39$ kHz). Additionally, we tuned with AROWS to the J$_{\rm{K_a,K_c}} =$ 2$_{1,1}$ e$_0$--2$_{0,2}$ e$_0$ (b-type) transition of CH$_2$DOH at 86.668751 GHz.  We observed Seo09, 12, and 17 in this transition during 3 shifts between 2018 November and December.

\begin{table}
	\centering
	\caption{Source core numbers and coordinates are from \citealt{2015ApJ...805..185S}. Column density of methanol taken from \citealt{2020ApJ...891...73S}.}
	\label{physparams}
	\begin{tabular}{cccc}
		\hline
		Core Number & $\alpha$ & $\delta$ & N$_{\rm{CH_3OH}}$ \\
		 & (J2000.0) & (J2000.0) & ($10^{13}$ cm$^{-2}$) \\
		\hline
		6 & 4:17:52.8 & $+$28:12:26 & $2.6 \pm 0.1$ \\
        7 & 4:18:02.6 & $+$28:10:28 & $1.0 \pm 0.1$ \\
        8 & 4:18:03.7 & $+$28:07:17 & $1.4 \pm 0.1$  \\
        9 & 4:18:07.0 & $+$28:05:13 & $2.3 \pm 0.1$ \\
        10 & 4:17:37.6 & $+$28:12:02 & $2.2 \pm 0.1$ \\
        11 & 4:17:51.2 & $+$28:14:25 & $2.6 \pm 0.1$ \\
        12 & 4:17:41.7 & $+$28:08:46 & $2.9 \pm 0.1$ \\
        13 & 4:17:42.2 & $+$28:07:29 & $2.5 \pm 0.1$ \\
        14 & 4:17:43.0 & $+$28:06:00 & $3.4 \pm 0.2$ \\
        15 & 4:17:41.1 & $+$28:03:50 & $1.5 \pm 0.1$ \\
        16 & 4:17:36.1 & $+$28:02:57 & $1.6 \pm 0.1$ \\
        17 & 4:17:50.3 & $+$27:55:52 & $0.8 \pm 0.1$ \\
	\end{tabular}
\end{table}

Observations were conducted in one 8-hour shift per source, with 5 minute absolute position-switching scans between the source and off position every 30 seconds. The off position was the same position as used for NH$_3$ observations in \cite{2015ApJ...805..185S}: 4$^h$18$^m$36.3$^s$ +28$^o$5$^{\prime}$43.6$^{\prime\prime}$ J2000.0. Observed temperatures were converted from the antenna temperature scale T$_\textrm{A}^*$ to the main beam temperature  T$_\textrm{mb}$ using measured beam efficiencies $\eta_\textrm{mb}$ \citep{1993PASP..105..117M} correcting for the sideband rejection (typically better than 15 dB). Beam efficiencies were calculated from continuum observations of Mars, averaged over the two linear polarizations. These averaged beam efficiencies were $\eta_\textrm{mb}$ = 0.830 $\pm$ 0.026.
There is an additional calibration ratio of 1.14 in T$_\textrm{A}^*$ between newer AROWS spectrometer data and older MAC spectrometer data that was multiplied into all MAC observations.
The average baseline RMS was $\sigma_{T_\textrm{mb}} = 11$ mK. 

Sources in the B10 regions are at an average distance of 135 pc \citep{2014ApJ...786...29S} with an average angular size of about 1$^{\prime}$ that is well-matched to our resolution. The FWHM beam size of 67$^{\prime\prime}$ at this distance corresponds to 0.044 pc.

\section{Data Reduction and Analysis}

\begin{figure*}
    \centering
    \includegraphics[width=176mm]{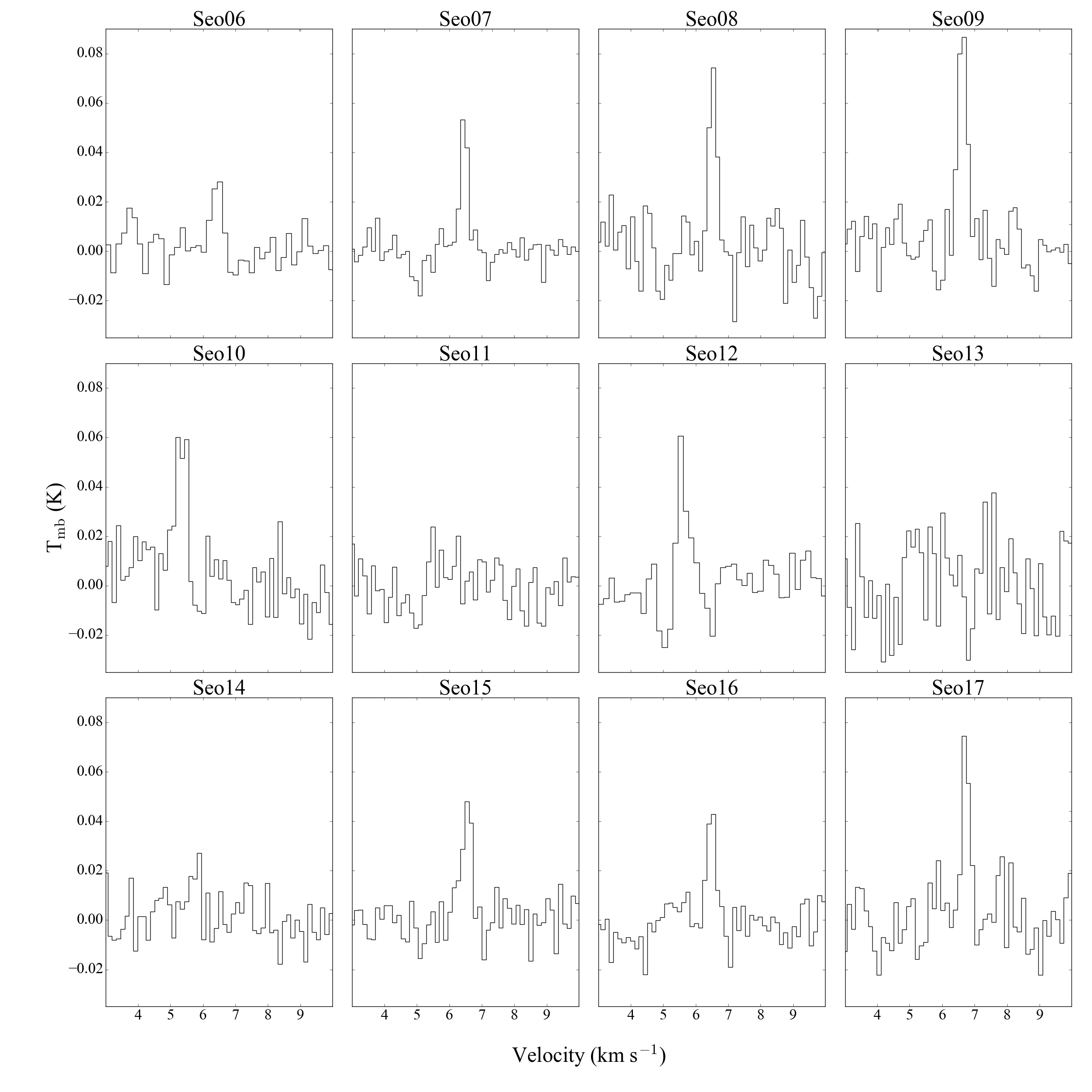}
    \caption{Spectra of all observed sources for the 89 GHz 2$_{0,2}$ e$_0$--1$_{0,1}$ e$_0$ (a-type) transition of CH$_2$DOH. Detection peaks can be seen for all sources except Seo 11, 13, and 14.  The velocity scale is radio $\mathrm{v}_{\rm{LSR}}$.}  
    \label{fig:gridofspectra}
\end{figure*}

CH$_2$DOH was detected in 9 of the 12 observed cores with greater than $3\sigma_{\textrm T_\textrm{mb}}$ certainty. Non-detections were taken to have an intensity upper-limit of 3 times the integrated intensity $rms$, which averaged $3\sigma_\textrm{I} = 0.017$ K km s$^{-1}$.
These data were reduced using the CLASS program of the GILDAS package\footnote{https://www.iram.fr/IRAMFR/GILDAS/}, with which Gaussian line profiles were fit to the spectra to calculate integrated intensity, velocity, and FWHM linewidth (see Table \ref{derivedparams}) using standard CLASS routines (\citealt{2005sf2a.conf..721P}, \citealt{2013ascl.soft05010G}). The symmetric, Gaussian shapes of the spectra were well-fit by a single Gaussian profile (Fig. \ref{fig:gridofspectra}).
All CH$_2$DOH linewidths are narrower than the corresponding CH$_3$OH linewidth with a median linewidth ratio of 0.78 (Table \ref{derivedparams}).

The deuterium fraction, f$_\textrm D$ = N(CH$_2$DOH)/N(CH$_3$OH), of methanol in the cores is calculated by deriving the column density of CH$_2$DOH and dividing by the total column density (adding A + E symmetry states) of CH$_3$OH previously published by \cite{2020ApJ...891...73S}.  CH$_3$OH column densities (see Table \ref{physparams}) were determined using the one dimensional non-LTE RADEX code \citep{2007A&A...468..627V}.  Three transitions of CH$_3$OH were modeled assuming the core was characterized by a homogeneous average density derived from \textit{Herschel} continuum observations \citep{2013A&A...550A..38P, 2016MNRAS.459..342M} and an isothermal gas kinetic temperature derived from NH$_3$ observations \citep{2015ApJ...805..185S} with no large scale velocity motions (see \cite{2020ApJ...891...73S} for details).  Since there are no published collision rates for CH$_2$DOH, we cannot use RADEX to estimate column densities for the deuterated species.
Instead, the column density of CH$_2$DOH (cm$^{-2}$) is determined from
\begin{equation}
    N(\rm{CH}_2\rm{DOH}) =  \frac{4\pi}{\rm{h c g_u A_{ul}}} \frac{B_\nu(T_{ex})Q(T_{ex}) e^{E_u / kT_{ex}}}{[J_\nu(T_{ex}) - J_\nu(T_{cmb})]} I(\rm{CH}_2\rm{DOH})
\end{equation}
where $\rm{A_{ul}}$ is the spontaneous emission coefficient of the CH$_2$DOH transition at frequency $\nu$, $\rm{E_u}$ and $\rm{g_u}$ are the upper level energy and statistical weight, T$_\textrm{ex}$ is the excitation temperature of the CH$_2$DOH transition, $Q$ is the partition function, I(CH$_2$DOH) is the integrated intensity of the line (K cm s$^{-1}$), and $B_\nu$ and $J_\nu$ are Planck functions in intensity and temperature units, respectively (see \citealt{2015PASP..127..266M} for a derivation). 
For the 89 GHz a-type transition, we assumed values derived from the JPL line catalog entry\footnote{https://spec.jpl.nasa.gov/ftp/pub/catalog/c033004.cat} of A$_\textrm{ul}$\,=\,2.02\,$\times 10^{-6}$ s$^{-1}$, g$_\textrm u$ = 5, and E$_{\rm{u}}$/k = 6.44 K.  For the 86 GHz b-type transition, we used A$_\textrm{ul}$ = 4.65 $\times 10^{-6}$ s$^{-1}$, g$_\textrm u$ = 5, and E$_{\rm{u}}$/k\,=\,10.60 K.
This equation assumes that the CH$_2$DOH emission is optically thin, that the filling fraction of emission is equal to 1, and that all of the rotational energy levels are populated with the same excitation temperature (the CTEX or constant excitation temperature approximation; \citealt{2002ApJ...565..344C, 2015PASP..127..266M}).  Given the different energy level structure of CH$_2$DOH compared to CH$_3$OH, it is likely that the CH$_2$DOH excitation temperature is different than the excitation temperatures calculated by RADEX for CH$_3$OH transitions.

\begin{figure*}
    \centering
    \includegraphics[width=140mm]{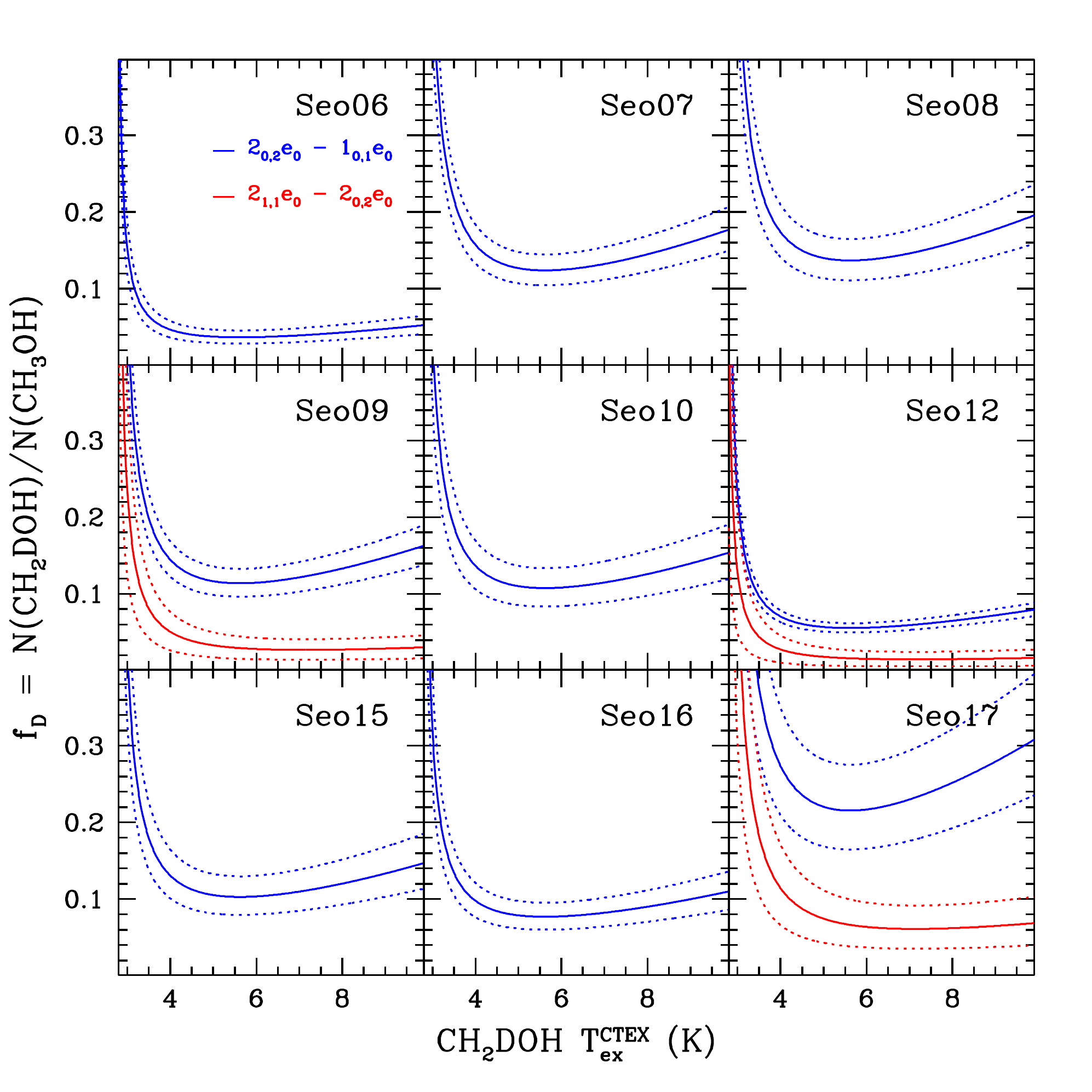}
    \caption{The deuterium fraction f$_\textrm D =$ N(CH$_2$DOH)/N(CH$_3$OH) is plotted for the detected sources versus the excitation temperature (CTEX approximation) of the CH$_2$DOH transition.  The blue curves correspond to the a-type $2_{0,2}$e$_0$--$1_{0,1}$e$_0$ transition and the red curves correspond to the b-type $2_{1,1}$e$_0$--$2_{0,2}$e$_0$ transition.  The solid line is f$_\textrm D$, while the dashed lines are $\pm 1\sigma$.}
    \label{fig:fdcurves}
\end{figure*}

Typically, observations of two optically thin transitions with different E$_{\textrm{u}}$/k at similar spatial resolutions may be used to simultaneously constrain the total column density and excitation temperature in the CTEX approximation; however, the uncertainty associated with calculating the matrix elements for b-type and c-type transitions of CH$_2$DOH presents a challenge with our observations (see Appendix \ref{appendA1}). 
We attempted to overlay the column density versus T$_\textrm{ex}$ curves using the same  procedure outlined in \citealt{2020ApJ...891...73S} for the three sources for which two CH$_2$DOH transitions were observed in this survey, but the curves did not overlap for T$_\textrm{ex} >$ T$_\textrm{cmb}$ (Fig. \ref{fig:fdcurves}).  This either indicates that there is significant non-LTE excitation occurring for the CH$_2$DOH  transitions (with population inversions) in prestellar cores or that we do not have an accurate value for the Einstein A of the b-type transition to make a reasonable estimate of T$_\textrm{ex}$ using this method (see Appendix \ref{appendA1}).

Since a reliable single value for the CH$_2$DOH T$_\textrm{ex}$ could not be calculated from the two observed transitions, the column densities of CH$_2$DOH were calculated for a range of possible excitation temperatures using only the a-type transition. 
These column densities were divided by the total CH$_3$OH column densities (adding A + E symmetry states) measured for the cores \citep{2020ApJ...891...73S} to obtain values of f$_\textrm D$.
A plot of f$_\textrm D$ vs T$_\textrm{ex}$ of CH$_2$DOH for each detected source shows that the resultant curves increase for T$_\textrm{ex} < 4$ K, and climb again for T$_\textrm {ex} > 8$ K (Fig. \ref{fig:fdcurves}).  Between these values, however, f$_\textrm D$ is relatively flat.
As f$_\textrm D$ is slightly higher for all sources at T$_\textrm{ex}$=4 K than T$_\textrm{ex}$=8 K, we took 4 K as our excitation temperature lower limit, and therefore as an upper limit on the column density of CH$_2$DOH and f$_\textrm D$ for each source.  Admittedly, this exact choice is arbitrary; however, the CH$_3$OH excitation temperatures of these sources typically range from $\sim$ 7--8 K \citep{2020ApJ...891...73S}.
Given the large deuteration values we find below, it therefore seems unlikely that CH$_2$DOH has excitation temperatures much below 4 K. 
The curves 
for all sources
reach a minimum f$_\textrm D$ at T$_\textrm{ex}$ = 5.6 K, and we took that T$_\textrm{ex}$ value for the 
column density of CH$_2$DOH and
f$_\textrm D$ lower limits. We took the median values between
the lower and upper limits described to estimate the column density of CH$_2$DOH and f$_\textrm D$ for each core. f$_\textrm D$ was found to range between 0.04 and 0.23 for the detected sources (Table \ref{derivedparams}).

\begin{table*}
	\centering
	\caption{CH$_2$DOH Observations of starless and prestellar cores in B10. Spectra are analyzed on the main beam temperature scale.  All uncertainties for detected qunatities are 1$\sigma$. Non-detections for I and f$_\textrm D$ values are reported as 3$\sigma$ upper limits. *Seo06 and Seo12 are observed with a 0.0819 km s$^{-1}$ resolution. All other cores have a resolution of 0.131 km s$^{-1}$}
	\label{derivedparams}
	\begin{tabular}{ccccccc}
		\hline
		Transition & Core Number & v$_{\rm{LSR}}$ & $\Delta$v  & I  & N$_{\rm{CH_2DOH}}$  & f$_\textrm D$\\
		 & & (km s$^{-1}$) & (km s$^{-1}$) & (mK km s$^{-1}$) & ($10^{12}$ cm$^{-2}$) & \\
		\hline
		& 6* & 6.43 $\pm$ 0.03 & 0.34 $\pm$ 0.07 & 11.8 $\pm$ 0.2 & 1.0$^{+0.4}_{-0.2}$ & 0.04$^{+0.02}_{-0.01}$\\
        & 7 & 6.45 $\pm$ 0.01 & 0.26 $\pm$ 0.03 & 15.9 $\pm$ 0.2 & 1.3$^{+0.5}_{-0.2}$ & 0.14$^{+0.06}_{-0.03}$\\
        & 8 & 6.54 $\pm$ 0.02 & 0.29 $\pm$ 0.04 & 23.3 $\pm$ 0.4 & 2.0$^{+0.8}_{-0.4}$ & 0.14$^{+0.07}_{-0.03}$\\
        & 9 & 6.62 $\pm$ 0.02 & 0.34 $\pm$ 0.04 & 33.7 $\pm$ 0.4 & 2.9$^{+1.0}_{-0.5}$ & 0.13$^{+0.05}_{-0.02}$\\
        & 10 & 5.32 $\pm$ 0.03 & 0.43 $\pm$ 0.07 & 29.2 $\pm$ 0.6 & 2.5$^{+1.1}_{-0.6}$ & 0.11$^{+0.06}_{-0.03}$\\
        2$_{0,2}$e$_0$ - 1$_{0,1}$e$_0$ & 11 & ... & ... & $<$13.3 & $<$1.4 & $<$0.06 \\
        & 12* & 5.56 $\pm$ 0.02 & 0.22 $\pm$ 0.04 & 19.2 $\pm$ 0.2 & 1.6$^{+0.5}_{-0.2}$ & 0.06$^{+0.02}_{-0.01}$\\
        & 13 & ... & ... & $<$24.6 & $<$2.5 & $<$0.11 \\
        & 14 & ... & ... & $<$12.2 & $<$1.3 & $<$0.04 \\
        & 15 & 6.54 $\pm$ 0.03 & 0.36 $\pm$ 0.07 & 18.5 $\pm$ 0.3 & 1.6$^{+0.7}_{-0.3}$ & 0.11$^{+0.06}_{-0.03}$\\
        & 16 & 6.48 $\pm$ 0.02 & 0.30 $\pm$ 0.05 & 14.8 $\pm$ 0.3 & 1.3$^{+0.6}_{-0.3}$ & 0.08$^{+0.04}_{-0.02}$\\
        & 17 & 6.72 $\pm$ 0.02 & 0.27 $\pm$ 0.05 & 22.3 $\pm$ 0.4 & 1.9$^{+0.8}_{-0.4}$ & 0.23$^{+0.12}_{-0.06}$\\
        \hline
        & 9 & 6.77 $\pm$ 0.04 & 0.38 $\pm$ 0.08 & 10.4 $\pm$ 2.3 & ... & ... \\
        2$_{1,1}$e$_0$ - 2$_{0,2}$e$_0$ & 12 & 5.58 $\pm$ 0.03 & 0.25 $\pm$ 0.09 & 7.2 $\pm$ 2.1 & ... & ... \\
        & 17 & 6.73 $\pm$ 0.02 & 0.23 $\pm$ 0.04 & 17.3 $\pm$ 8.5 & ... & ... \\
		\hline
	\end{tabular}
\end{table*}

\section{Discussion}

\subsection{Comparison with Chemical Models}

Our observed abundance ratios are consistent with observations of L1544, which showed a deuterium fractionation of methanol of f$_\textrm D$= 0.08 $\pm$ 0.02 \citep{2017MmSAI..88..724C,2019A&A...622A.141C}. Because f$_\textrm D$ does not vary strongly with T$_{\textrm{ex}}$ within the range of 4--8 K (typically 20 percent), our method of choosing T$_{\textrm{ex}}$ for this paper is consistent with the methods used by \cite{2014A&A...569A..27B} and \cite{2019A&A...622A.141C} to estimate the excitation temperature of singly-deuterated methanol.
Since observations are limited to gas-phase molecules, the abundance on the grain surfaces is unknown; therefore, we must compare our observations to predictions of gas-grain chemical models.

Methanol is predicted to be deuterated on the icy grain surfaces either by the addition of atomic D to CO
, by deuterium-substitution reactions with methanol, or by a combination of the two \citep{1983A&A...119..177T,1997ApJ...482L.203C,2012ApJ...748L...3T}.
The first process requires 
CO freezeout to begin, as well as high atomic D/H ratios \citep{2002P&SS...50.1257C, 2006A&A...453..949P}. 
Large gas phase atomic D/H ratios ($> 0.1$) are predicted in regions where significant CO freezout has occurred as a result of the dissociative recombination of fractionated H$_2$D$^+$, D$_2$H$^+$, and D$_3^+$ \citep{2003ApJ...591L..41R}.
Indeed, the D/H ratio has been measured to increase in prestellar cores for higher CO depletion, and thus higher density \citep{2002A&A...389L...6B, 2003ApJ...585L..55B, 2005ApJ...619..379C}.
The process of methanol deuteration via deuterium-substitution reactions has been measured in the laboratory and can reproduce large deuterium fractions \citep{2005ApJ...624L..29N}.

\cite{2012ApJ...748L...3T} analyzed the deuteration of methanol on the dust grains of late-stage, low-mass prestellar cores using the GRAINOBLE model introduced in \citealt{2012A&A...538A..42T}. GRAINOBLE is a time-dependent gas-grain chemical model which relies on the rate equations method, differing from previous models by treating the grain as a reactive outer layer surrounding an inert mantle bulk \citep{1993MNRAS.263..589H,2011ApJ...735...15G} while including both H and D addition to atomic CO and H/D substitution reactions, and allowing the H/D ratio to vary throughout the simulation \citep{2002P&SS...50.1257C,2003MNRAS.340..983S}.
The model finds that for $n_H= 10^5$ cm$^{-3}$, similar to the average density of cores dervied by \cite{2020ApJ...891...73S},
the mean $f_D =$ [CH$_2$DOH]/[CH$_3$OH] in the ice reaches values ranging from 0.05--0.16 by t = $10^6$ yr.
These model estimates agree with the median values of f$_\textrm D$ observed in the gas phase towards the B10 region for all cores detected in methanol with the exception of Seo17, whose lower limit value (f$_D$=0.165) is just above the model range. 
This would indicate, if the model predictions are accurate, that a significant fraction of the CH$_2$DOH formed in the ices at densities of $10^5$ cm$^{-3}$ desorbes into the gas phase.

For higher density values ($\sim 5 \times 10^{6}$ cm$^{-3}$), which will occur as the prestellar core evolves towards a first hydrostatic core, the \cite{2012ApJ...748L...3T} models predict even higher ice deuterium fractions.  This should be observable during the initial stages of protostellar accretion as heating desorbs the ice mantles. Single-dish observations out to $\sim$ 1000 AU by \cite{2002A&A...393L..49P,2004A&A...416..159P,2006A&A...453..949P} indeed show higher deuterium fractions ($\sim$ 0.3--0.6) for methanol in Class 0 protostars. 
Ultimately, the deuterium fraction of methanol is predicted to decrease towards the warm central regions ($<$ 100 AU) of the core envelope as the protostar evolves \citep{2014ApJ...791....1T}, which has been confirmed by recent interferometric observations of Class 0/I protostars \citep{2017MNRAS.467.3011B, 2017A&A...606L...7B,2018A&A...620A.170J,2018A&A...610A..54P,2019A&A...632A..19T,2020A&A...635A..48M, 2020arXiv200506784V}.

\subsection{Evolutionary Comparisons}

\begin{figure}
    \centering
    \includegraphics[width=83mm]{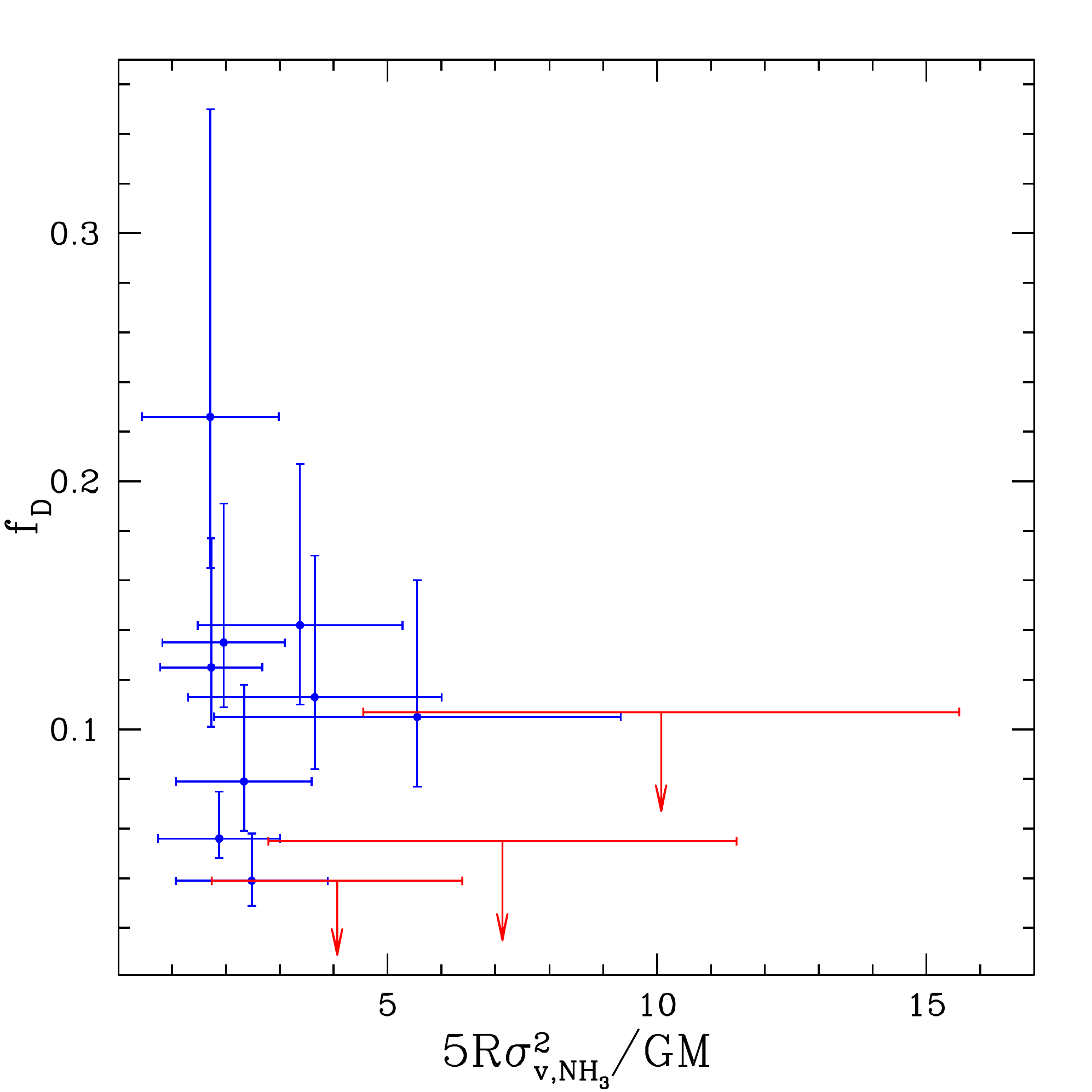}
    \includegraphics[width=83mm]{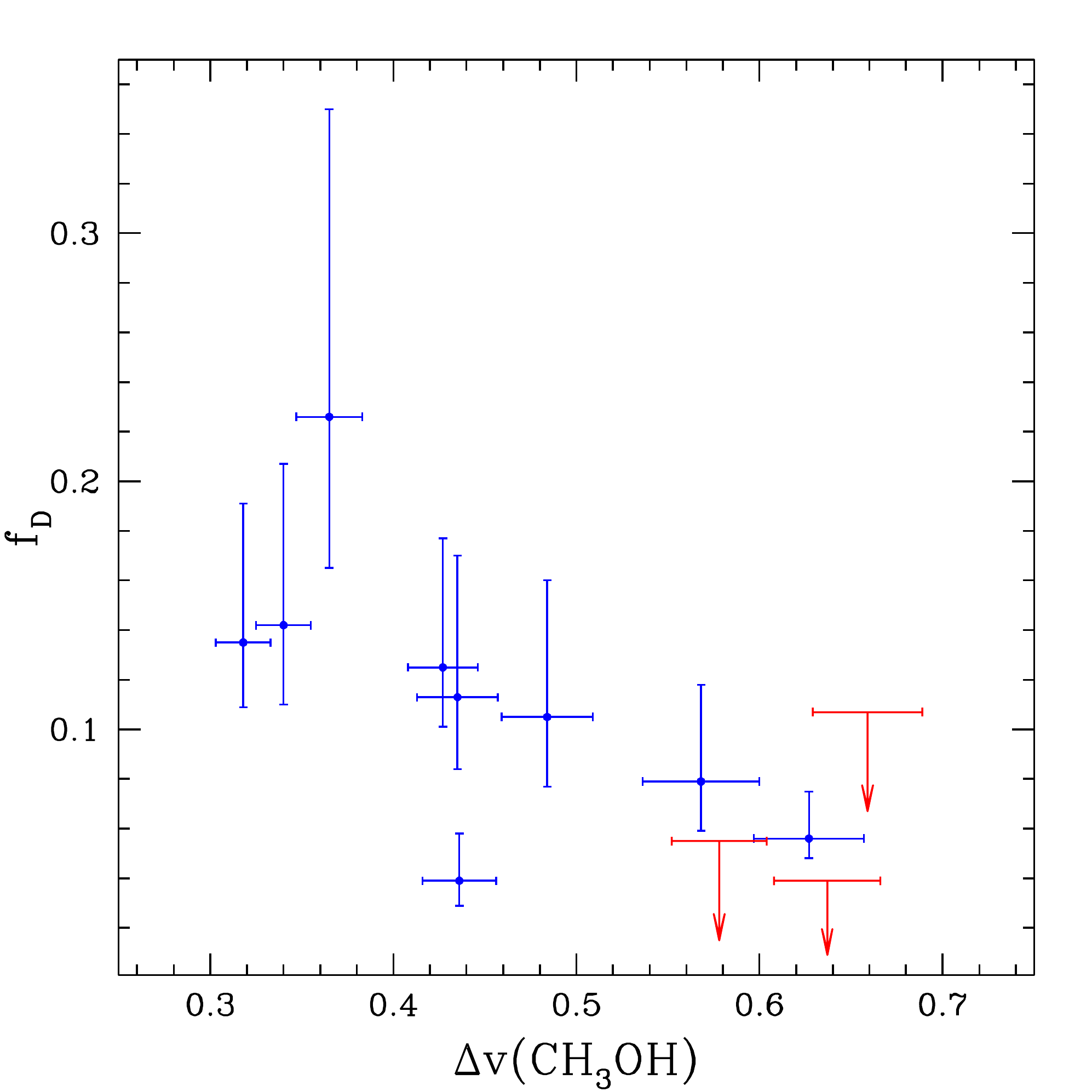}
    \caption{(Top) A plot of deuterium fraction f$_\textrm D =$ N(CH$_2$DOH)/N(CH$_3$OH) vs. the virial ratio $\alpha = |\frac{2 \Omega_\textrm K}{\Omega_\textrm U}|$ and (Bottom) linewidth, $\Delta$v, of CH$_3$OH-A J$_{\rm{K}} = $ $2_0 - 1_0$  for the starless cores of the B10 region.  Sources detected in CH$_2$DOH are shown in blue while non-detections are shown in red with $3\sigma$ upper limits.}
    \label{fD}
\end{figure}

Deuterium fractionation has traditionally been used as a chemical evolutionary indicator towards starless and prestellar cores \citep{2005ApJ...619..379C}.  Some molecules, such as NH$_3$ or N$_2$H$^+$, are considered "late-time" species since their abundances tend to peak during more advanced stages of starless core evolution.  In constrast, CH$_3$OH is considered an "early-time" species since its abundance peaks early in the core's evolution when CO begins to deplete on to dust grains, after which its abundance decreases as CH$_3$OH itself freezes out of the gas phase in the central dense, cold regions.  CH$_2$DOH is a probe of deuterium fractionation in an "early-time" species; however, there is difficulty in assigning CH$_2$DOH as an "early-time" species since deuterium fraction tends to increase during starless core evolution.  
There is also difficulty in assigning CH$_2$DOH as a "late-time" species, as maps of L1544 indicate that emission from this molecule does not peak at the dust continuum peak as other "late-time" species do \citep{2019A&A...622A.141C}.
With this tension in mind, we have investigated how the methanol deuterium fraction varies with evolutionary parameters of the cores (Fig. \,\ref{fD}). 

As a starless core evolves to a prestellar core and eventually to a first hydrostatic core, the central density of the core increases and the core becomes more gravitationally bound. 
The virial ratio is one metric for determining the evolutionary stage of starless cores. The virial ratio $\alpha$ is defined as the ratio of twice the total kinetic energy to the total gravitational potential, $\alpha = |\frac{2\Omega_\textrm K}{\Omega_\textrm G}|$.  
Since starless cores have density profiles that tend to be flat towards the centre and then fall off like a power-law (i.e. Bonnor-Ebert Spheres or Plummer-like Spheres; \citealt{1956MNRAS.116..351B, 1955ZA.....37..217E, 2001ApJ...547..317W}) and since the change of the numerical coefficient in the gravitational potential energy term between a uniform density sphere and a critical Bonnor-Ebert Sphere is only 0.6 to 0.73 respectively \citep{2011A&A...535A..49S}, we shall simply assume the coefficient for a uniform density core and therefore $\alpha = 5 R \sigma_v^2/$GM.
Larger values of $\alpha$ indicate less gravitationally-bound objects and smaller values indicate more gravitationally-bound (and presumably more evolved) objects. 
We note that this simple virial ratio ignores potentially important contributions from the external pressure, internal magentic fields, and any mass flow across the core boundary; however, it simply provides a quick estimate of the importance of gravitational potential energy to the internal kinetic energy.
Virial ratios for the cores in the survey were reported in \cite{2020ApJ...891...73S} and were calculated using the size of the NH$_3$ core dendrogram, the NH$_3$ (1,1) velocity dispersion \citep{2015ApJ...805..185S}, and the mass within the core dendrogram boundary determined from the \textit{Herschel}-derived N(H$_2$) map of \cite{2013A&A...550A..38P}. 
The uncertainty on the dendrogram-derived core size is assumed to be one half of a pixel around the boundary. Errorbars on the virial ratio are statistical and do not include systematic uncertainties, for instance, in the assumed dust opacities \citep{2005ApJ...632..982S, 2011ApJ...728..143S, 2019MNRAS.489..962H}.
A plot of f$_\textrm D$ vs. $\alpha$ (Fig. \,\ref{fD}) shows that cores not detected in CH$_2$DOH tend to have higher virial ratios ($\alpha > 4$), although the statistical errorbars on the virial ratio are substantial.  
This result is consistent with CH$_2$DOH non-detections towards less-evolved starless cores. 
There also is no significant correlation between CH$_3$OH deuterium fraction and the virial parameter.

The most robust trend (Pearson correlation coefficient of $-0.63$ and rank correlation coefficent of $-0.82$) is that the deuterium fractionation is anti-correlated with the FWHM linewidth of the CH$_3$OH molecule (Fig\,\ref{fD}).  The FWHM linewidth is a result of the gas motions within the telescope beam, with more turbulent sources exhibiting larger non-thermal doppler motions, and therefore larger linewidths. 
We note that CH$_3$OH is not exclusively a dense core tracer, as extended CH$_3$OH emission is seen in the lower density filamentary structure in which the starless cores are embedded  \citep{2020ApJ...891...73S}.
Since the turbulent motions of starless cores dissipate as the cores evolve, the linewidths of their spectra decreases with age.  
Three of the four starless cores with the largest linewidths are non-detections in CH$_2$DOH.
The ratio of the linewidths of non-detections to the linewidths of detections is 1.4.
This result is consistent with the conclusion that higher deuterium fraction is associated with more-evolved cores.

\section{Conclusions}

Observations of twelve starless and prestellar cores in the B10 region of the Taurus Molecular Cloud show the presence of gas phase singly-deuterated methanol, CH$_2$DOH, in 75 percent of its prestellar dense cores.
All CH$_2$DOH FWHM linewidths are narrower than the corresponding CH$_3$OH linewidths with a median ratio of 0.78.
The deuterium fraction
of CH$_2$DOH to its non-deuterated form, CH$_3$OH, ranges from $< 0.04$ to 0.23$^{+0.12}_{-0.06}$ with a median value of $0.11$.
These values are consistent with previous observations of the prestellar dense core L1544, also in Taurus, 
as well as with the theoretical values derived in the \cite{2012ApJ...748L...3T} ice chemistry model.
The cores detected in CH$_2$DOH appear to be more-evolved, on average, than their non-detected counterparts (which have  larger virial parameters and larger CH$_3$OH linewidth than detected cores), though the region as a whole is relatively young compared to other regions within the Taurus Molecular Cloud. This survey indicates that deuterium fractionation in  organic molecules such as CH$_3$OH may be more easily detectable than previously thought and, in the case of CH$_2$DOH, may provide a probe of the deuteration on the icy surfaces of dust grains from molecules which are desorbed into the gas phase. Our observations provide new constraints for gas-grain astrochemical models of starless and prestellar cores that are more representative of a typical core than some of the more extreme objects studied to date.

\section*{Acknowledgements}

We sincerely thank the referee for their comments which improved this paper.  We also sincerely thank the staff and the operators of the Arizona Radio Observatory (Michael Begam, Kevin Bays, Robert Thompson, and Clayton Kyle) for their assistance with the observations. 
We are thankful that we have the opportunity to conduct astronomical research on Iolkam Du'ag within the Tohono O'odham Nation.
Hannah Ambrose, Yancy Shirley, and Samantha Scibelli were supported by NSF Grant AST-1410190 (PI Shirley). 
Samantha Scibelli is also supported by a
National Science Foundation Graduate Research Fellowship (NSF GRF) Grant DGE-1143953.
The 12 m Telescope is operated by the Arizona Radio Observatory (ARO), Steward Observatory, University of Arizona, with funding from the State of Arizona, NSF MRI Grant AST-1531366 (PI Ziurys), NSF MSIP grant SV5-85009/AST- 1440254 (PI Marrone), NSF CAREER grant AST-1653228 (PI Marrone), and a PIRE grant OISE-1743747 (PI Psaltis).

\section*{Data Availability}

The data underlying this article will be shared on reasonable request to the corresponding authors.


\bibliographystyle{mnras}
\bibliography{biblio.bib}



\appendix
\section{Rotational Transitions of Singly-Deuterated Methanol} \label{appendA1}

\begin{figure*}
    \centering
    \includegraphics[width=115mm]{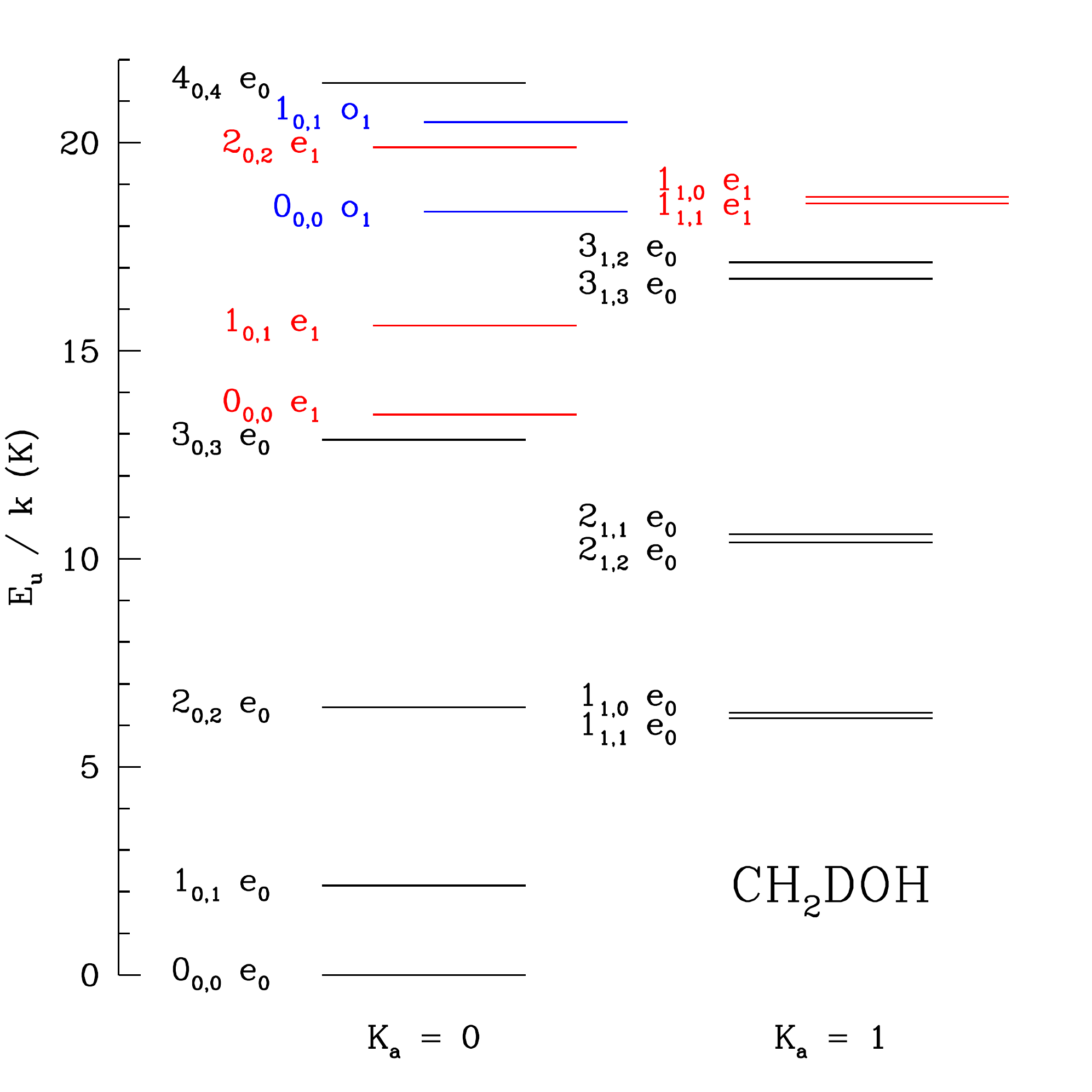}
    \caption{Energy level diagram for bound states of CH$_2$DOH with $\mathrm{E_u/k} < 22$ K.  Different torsional symmetry substates are plotted as slightly offset within each $\mathrm{K_a}$ ladder and color coded as $\rm{e}_0$ (black), $\rm{e}_1$ (red), and $\rm{o}_1$ (blue).  Each torsional symmetry substate follows the same general pattern of asymmetric top rotational energy levels but offset in energy from each other. Note that the lowest energy level with $\rm{K_a} = 2$ is the $2_{2,1}$ $\rm{e}_0$ level with $\mathrm{E_u/k} = 22.6$ K.}
    \label{fig:englvls}
\end{figure*}

CH$_2$DOH is an asymmetric top molecule with torsional motions that has allowed a-type transitions ($\Delta K_a = 0$, $\Delta K_c = \pm 1, \pm 3, ...$), b-type transitions ($\Delta K_a = \pm 1, \pm 3, ...$, $\Delta K_c = \pm 1, \pm 3, ...$), and c-type transitions ($\Delta K_a = \pm 1, \pm 3, ...$, $\Delta K_c = 0$).  The lowest rotational energy levels are shown in Figure \ref{fig:englvls}.
The methyl-D$_1$ (CH$_2$D) group is an example of an asymmetric hindered rotor in the CH$_2$DOH molecule \citep{2012JMoSp.280..119P, 2016InPhT..77..283M, 2016InPhT..79..216M} that results in 3 energetically-separated torsional symmetry substates (even or odd functions of the torsional angle) for each ground state asymmetric top rotational level (J$_{\rm{K_a,K_c}}$) that are labelled as e$_0$ (trans), e$_1$ (gauche), and o$_1$ (gauche).  Since the molecular a-axis component of the dipole moment is independent of the torsional angle, a-type transitions are only permitted within the same torsional symmetry substate (i.e. e$_0 \rightarrow$ e$_0$) and their matrix elements are more precisely calculated.
For b-type transitions, 
the torsional substate selection rules are more complicated and can change substates while maintaining torsional substate parity 
(i.e., $e_0 \rightarrow$ e$_0$, $e_1 \rightarrow$ e$_1$, e$_0 \leftrightarrow$ e$_1$, and o$_1 \rightarrow$ o$_1$ allowed). 
c-type transitions cross torsional substates with different symmetry 
(i.e., e$_0  \leftrightarrow$ o$_1$ allowed). 
The matrix elements for b-type and c-type transitions are more uncertain than for a-type transitions as they involve the mutliplication of the torsional overlap integral (a function of torsional excitation, substate and symmetric top K value which is a strong function of J) times the rotational dipole matrix element, and the b and c dipoles have significant 3-fold sine and cosine terms that are unknown (J. Pearson 2019, private communication).
The JPL line catalog entry for CH$_2$DOH has a dire warning to use extreme caution when determining column densities from b-type and c-type transitions\footnote{https://spec.jpl.nasa.gov/ftp/pub/catalog/doc/d033004.pdf}.  Our analysis in this paper confirms this warning.


\bsp
\label{lastpage}
\end{document}